# Thermomechanical properties of polypropylene and styrene-ethylene-butylene-styrene blends: a molecular simulation and experimental study

Pashupati Pokharel[a], Feng Wei[b], Jianyi Shi[b], Yingmin Wang[b], and Dequan Xiao[a]*

[a]Center for Integrative Materials Discovery, Department of Chemistry and Chemical Engineering, University of New Haven, Boston Post Road, West Haven 06516, Connecticut, USA

[b] Higasket Technical Center, Hefei National Economic-Technological Development Area Hefei, 230601

**Abstract**

Polymer blends consisting of two or more polymers are important for a wide variety of industries and processes, but, the precise mechanism of their thermomechanical behaviour is incompletely understood. In order to understand clearly, it is essential to determine the miscibility and interactions between the components in polymer blend and its macroscopic thermomechanical properties. In this study, we performed experiments on SEBS and isotactic PP blends (SP) as well as molecular dynamics simulations, aiming to know the role played by molecular interactions on the thermomechanical properties. To investigate the glass transition temperature ($T_g$) of SEBS, PP and their blends at different ratio, the unit cell of the polymer molecular structure of each was established. The LAMMPS molecular dynamics method was used to predict the density, specific volume, free volume, enthalpy, kinetic energy, potential energy and bond energy. The ($T_g$) s of the SEBS, PP and SP blends were predicted by analysing these properties. Interestingly, the simulated values of the $T_g$ of SEBS, PP and their blends showed good agreement with our experimental results obtained from dynamic mechanical analysis (DMA). This technique used in this work can be used in studying glass transition of other complex polymer blends.

**Introduction**

Blending and co-polymerization offered a pair of versatile and cost-effective procedures from combinations of existing chemicals for the production of numerous chemical having superior assets, mechanical properties (dimensional stability, abrasion resistance, impact strength, fracture toughness, etc.), thermophysical properties (e.g., thermal stability, melting point, degree of crystallinity, and crystallization rate), electrical or dielectric properties (e.g., conductivity and permittivity), etc. effortlessly [1–8]. Thermal events, such as a glass transition i.e. the thermal softening point of the polymer is important in selecting materials for end use applications. Glass transition of polymer denotes a region of the reversible transformation of amorphous materials (including amorphous regions within semi-crystalline polymers) from a rubber-like state into a stiff and relatively brittle glassy state [9–13]. The glass-transition temperature ($T_g$) is very reproducible and has become acknowledged as one of the most important material properties, directly relating to a number of other thermo-physical and rheological properties [9,11,14–17]. Nowadays, to achieve target for technological or industrial requirements in polymer engineering, it is largely beliefs on chemical and compositional manipulation of $T_g$. Quench cooling of molten solid polymers gets converted to amorphous solid form because of the dynamic arrest of molecules forming a disordered state at $T_g$. The polymer chains in glassy state are subject to only vibration and not translational and rotational motion. The conversion of glassy (amorphous) solid to rubbery (viscous liquid) takes place at $T_g$ in which numerous factors such as structural change in molecules, cooling rate and incorporation of additives alter the $T_g$ [12,18]. Techniques like differential scanning calorimetry and dynamic mechanical analysis are commonly used to measure the $T_g$ of polymers and polymer nanocomposites [14–16,19–23].

The chemical inertness, broad property window and cost competitiveness of polypropylene (PP) increased the popularity in applications which require softness and high transparency. To achieve the desired softness of the PP based final products, elastomers such as styrene-ethylene/butylene-styrene triblock-copolymers (SEBS) usually need to be added. Guo et al.[1] prepared the PP/PP-g-PS/SEBS blends by melt extrusion in order to improve both insulating properties and toughness of PP. They used SEBS to reduce the rigidity of PP, while the insulating properties are increased by adding PP-g-PS without diminishing the mechanical properties. Based on the evaluation of the microstructure of PP blends using SEM, DMA, XRD and DSC, they confirmed the formation of a core-shell dispersion phase in the blend with the adding of PP-g-PS, and the size of core is decreased while the thickness of shell is increased with further increasing volume of PP-g-PS. In their study, the rigid segments and compatibilization effect of PP-g-PS not only enhanced the glass transition temperature of both PP and SEBS, but also increased their adhesion. Pantoja et al. [24] showed the thermally responsive shape memory properties of neat SEBS and quantitatively investigated under uniaxial tension using a dynamic mechanical analyzer and manual stretching. They noticed the partial stress relaxation of the block copolymer network under load and the formation of a second network with a lower glass transition temperature, which are responsible for shape memory properties of SEBS. Martin et al. [2] demonstrated the intercalation capability of SEBS in nanocomposites of isotactic PP with 5 wt% of organically modified clay, prepared by melt blending. Based on the X-ray diffraction and electron microscopy, they proved that clay is not in direct contact with the PP phase because the clay is always located inside the elastomer domains. Furthermore, the elastomer is surrounding the nano-clay, obstructing the clay exfoliation and inhibiting its dispersion in the PP matrix.

Even though, there are lots of experimental study on SEBS and PP blends with and without nanofiilers and compatibilizer, there are not experimental and simulation approaches employed together to study the thermomechanical properties of PP and SEBS blend (SP) concerning the mutual effect of each component of the polymer blends yet. In SP blend, SEBS helps to achieve the desired film softness and toughness during blending with PP. We felt it is quite important to understand the proper ratio of SEBS and PP to minimize the cost with desired property profile with the reduction of costly SEBS in film-based packaging systems. Cast films prepared from SEBS and PP blends exhibit lower stiffness, higher toughness and improved optical properties compared to films prepared from the neat base resins [1,25]. Melt mixing provides one of the most common techniques for the large-scale preparation of polymer blends [3,9–13,26–28]. In case of analysis of polymer-based materials, $T_g$ is considered as a major property to modify their physical properties. Regarding the information of $T_g$, one can keep material in crystalline or amorphous state, viscous/rubbery/supercooled liquid and less viscous liquid form. Dynamic mechanical analysis (DMA) is an effective tool for the characterisation of viscoelastic materials [29]. Several researchers have investigated effects of blend ratio on the dynamic mechanical properties [30–34]. In this study, the dynamic mechanical properties such as storage modulus (E'), loss modulus (E'') and damping properties, Tanδ of blends of SEBS and PP was investigated with special reference to the effect of blend ratio over a temperature range -80°C to 160°C. The degree of polymer miscibility and damping characteristics in blends was studied based on the nature of Tanδ curves. The Tanδ curves of the SP blends exhibited three transition peaks at higher concentration of PP, corresponding to the glass transition temperature ($T_g$) of individual components, indicating incompatibility of the blend systems at high concentration of PP in SP blend.

Molecular dynamics (MD) simulation can be used as a tool to screen and design polymers and blends for end use applications. Simply, the segment motion of a polymer chain is used to predict the $T_g$ of polymeric materials. The polymer chain should have sufficient space and can complete this motion in a short time period [12,18]. However, the rigidity of a polymeric material determines the less or high relaxation time of polymer chains. The thermal softening point of the polymer i.e. $T_g$ is crucially important in aerospace applications [35].

Yang. et. al. [12] predicted the $T_g$ of the polyethylene unit cell by means of analysing density, free volume, specific volume, radial distribution function, non-bond energy, torsion energy, etc. of PE. Their simulated value of the $T_g$ is about 200 K, which is in good agreement with available data of 195 K in the literature. From the definition of $T_g$, the relaxation time scale at the $T_g$ is around 100 s. But MD simulations relate to much shorter time scales than the typical experiments nearly more than 10 orders of magnitude. However, based on Jie Han's reports, $T_g$ of polymeric materials from MD simulations performs to be a practical procedure [36]. Li et al. [37] applied molecular dynamics method to study the abnormal phenomenon of the $T_g$ of isomeric polyimide (PI) with its corresponding symmetrical PI. They predicted the chain rigidity of isomeric PI which is responsible for the higher $T_g$ than its corresponding symmetrical PI. There is not any literature of combining experimental measurements and MD simulation to study the $T_g$ of polymer blends specially prepared from mixing of elastomer and thermoplastic. Peng et al. [38] performed MD simulations on dynamically heterogeneous blend system containing soft (low-$T_g$) and stiff (high-$T_g$) polymer chains and having two different morphologies: homogeneous and phase-separated. At all shear frequencies simulated, they observed the storage modulus of the homogeneously mixed blends was greater than that of the phase-separated blends. Furthermore, in the homogeneous blend, the high-$T_g$ chains could slow down the dynamics of the low-$T_g$ matrix chains due to greater number of interactions between the two chain types in the well-mixed state. We felt an urgent to know the effect of addition of thermoplastic to the $T_g$ of hard segment and soft segment of elastomer for different industrial applications. In this study, experimental $T_g$ measurements of SEBS, PP and their blends at different molar ratio were performed to correlate with LAMMPS MD simulation results of polymer blends at different molar ratio. We believe that these MD simulations are especially useful for revealing the glass transition behaviour of complex polymeric blends.

## 2. Experimental

*2.1 Materials*

The two neat polymers chosen for the study were isotactic polypropylene (PP) and styrene-ethylene-butylene-styrene (SEBS) (Figure 1). PP and SEBS were purchased from Sigma Aldrich. The components of the PP/SEBS blends were physically mixed at two-roller mill at 180°C for 25 min. The molar mass ratio of SEBS and PP was fixed as shown in Table 1 for the preparation of the series of binary SP blends. All the chemical reagents were used without further purification.

*2.2 SP blend preparation*

SP blends with the variation of SEBS and isotactic PP weight ratio were prepared by melt mixing in two-roll mill (Laboratory roll mill, Fanyuan Instrument) with the variation between 5 to 30 rpm and a maximum temperature of 180°C. Thermal stabilizer (0.1 wt%) was added during melt processing to prevent the degradation of neat polymers and blends. Appropriate molar ratios of SEBS and PP (Table 1) were mixed at 180°C and (5-30) rpm for 25 min to obtain blends of different compositions. The blends are represented as SP1 (1:2), SP2 (1:1) and SP3 (2:1), where S stands for SEBS, P stands for PP and subscripts in bracket represent the molar weight ratio of SEBS and PP in the blend. Neat SEBS, neat PP as well as three SP blends were compression moulded in an electrically heated Hydraulic Hot Press (Carver Hydraulic Hot Press) at 165°C to get sheets of ~2 mm thickness for dynamic mechanical testing.

Table 1. Glass transition temperature ($T_g$) of neat SEBS, neat PP, SP1, SP2 and SP3 blends obtained from DMA and MD simulation.

| Sample code | Molar ratio | | $T_{g1}$ (Experimental) (K) | $T_{g1}$ (MD Simulation) (K) | $T_{g2}$ (Experimental) (K) | $T_{g2}$ (MD Simulation) (K) |
|---|---|---|---|---|---|---|
| | SEBS | PP | | | | |
| Neat SEBS | 1 | 0 | 231.5 | 229.8 | 386.27 | 385.8 |
| Neat PP | 0 | 1 | 282.3 | 282.1 | - | - |
| SP1 | 1 | 2 | 222.7 | 225 | - | 376.5 |
| SP2 | 1 | 1 | 216 | 215.1 | - | 380.0 |
| SP3 | 2 | 1 | 226.7 | 225.6 | 385.71 | 383.4 |

## 2.3. Characterization

The thin films of PP, SEBS, and binary SP blends having three different ratio of SEBS and PP were annealed at 70°C for 24 h and cut into small rectangular shape having breadth of ~6.5 mm and thickness ~2.0 mm for DMA analysis (dynamic mechanical analyser, TA instrument, DHR-2). The measurement was performed in tension mode, where the static force and dynamic force were taken as 10 and ± 5 N, respectively. The dynamic frequency was kept constant at 1.0 Hz and the heating rate was selected as 5°C/min from -80 to 160°C.

## 2.4 Simulation details

In molecular dynamics (MD) simulations, computation of microscopic interactions is used to describe the macroscopic behaviour of polymers. These microstates are created by exploring the both positions and velocities of the constituent atoms in the continuous space. The MD simulation considers all atoms as classical particles and shows the dynamic behaviour of molecules [39–41].

Based on Newton's second law of motion,

$$F = ma = m\frac{dv}{dt} = m\frac{d^2r}{dt^2} \quad\ldots\ldots\ldots\ldots\ldots\ldots.(1)$$

$$F = ma = \frac{\partial U(r)}{\partial r} \quad\ldots\ldots\ldots\ldots\ldots\ldots\ldots\ldots(2)$$

where F is force acting on each atom, m is mass of the particle, a is its acceleration and r is its position. Here, F is defined as a negative gradient of the potential energy function of the system which can be obtained from the potential energy, U(r) and is readily differentiable with respect to the atomic positions. For a given set of initial conditions, i.e. velocities and positions, equation 1 can be integrated numerically in distinct time steps for all the particles in the simulated system, to yield their positions and momenta as function of time. This permits calculation of the average values for the properties of concern [41].

*Force Fields*

Development of an accurate Force Field (FF) is quite important in atomistic simulations for use in molecular dynamics [40].

*The energy expression*

$$E_{total} = E_{valance} + E_{non\,bond} \quad\ldots\ldots\ldots\ldots\ldots\ldots\ldots\ldots.(3)$$

Where valence terms ($E_{valence}$) involves covalent bonds and long-range noncovalent interactions ($E_{nonbond}$).

The covalent terms can express as

$$E_{valance} = E_{bond} + E_{angle} + E_{torsion} \quad\ldots\ldots\ldots\ldots\ldots\ldots.(4)$$

It includes bond stretch ($E_{bond}$), angle bend ($E_{angle}$), and dihedral angle torsion ($E_{torsion}$).

The nonbond terms

$$E_{nonbond} = E_{vdW} + E_{Coulomb} \quad\ldots\ldots\ldots\ldots\ldots\ldots\ldots..(5)$$

It consists of van der Waals ($E_{vdW}$) and electrostatic ($E_{Coulomb}$) terms.

Valence interactions are described using bond stretch term, angle bend terms and torsion terms.

$$E_{bond}L = \frac{1}{2}K_L(L - L_0) \quad \ldots\ldots\ldots\ldots\ldots\ldots\ldots\ldots\ldots\ldots(6)$$

This expression is used to describe the bond stretch terms (Harmonic) where L is the bond length, $L_o$ is the equilibrium bond length, and $K_R$ is the force constant.

$$E_{angle\,\theta} = \frac{1}{2} K_\theta (\theta - \theta_0)^2 \quad \ldots\ldots\ldots\ldots\ldots\ldots\ldots(7)$$

Above expression is used to express angle bend terms where $\theta$ is the angle between two bonds and $\theta_0$ is the equilibrium length and $K_\theta$ is the force constant.

$$E_{torsion}(\emptyset) = \sum_{n=0}^{n=10} V_n \cos(n\emptyset) \ldots\ldots\ldots\ldots\ldots\ldots(8)$$

For the dihedral torsional angle ($\emptyset = 0$ corresponds to cis) and write the energy as a Fourier expansion as above.

*Nonbond Interactions*

Electrostatic terms and van der Waals Terms are commonly described under non-bond interaction.

*Electrostatic Terms (Coulomb)*

The electrostatic interactions between two atoms m and n is expressing using following equation

$$E_{Coulomb}(R_{mn}) = C_Q \frac{Q_m Q_n}{\varepsilon R_{ij}} \ldots\ldots\ldots\ldots\ldots\ldots\ldots\ldots\ldots(9)$$

where $Q_m$ is the charge on center m (electron units), $\varepsilon = 1$, and the constant $C_Q = 332.0637$ gives energies in kcal/mol and $R_{mn}$ is the distance in angstrom.

*van der Waals Terms*

The van der Waals interaction between atoms m and n are express as

$$E_{vdW}(L_{mn}) = D_v \left\{ \left(\frac{L_v}{L_{mn}}\right)^X - 2\left(\frac{L_v}{L_{mn}}\right)^Y \right\} \ldots\ldots\ldots\ldots\ldots\ldots(10)$$

where $L_{mn}$ is the distance between the atoms, $L_v$ is the equilibrium distance, $D_v$ is the well depth.

In this study, MD simulations were performed using MAPS 4.3 software purchased from Scienomics. The amorphous structures of neat SEBS, neat PP and SP blends unit cell were established in this work. Dreiding force field was employed for MD simulations [42–44]. The LAMMPS MD simulation was accomplished at constant volume constant-temperature (NVT) and constant pressure constant-temperature (NPT) ensemble. In this procedure, density of neat polymers as well as blends system was above 0.8 g/cm³. The system was considered equilibrate based on the stable thermodynamic quantities during the last 500 ps. Geometry optimization of the unit cells of neat polymers and their blends was performed at 450K by smart minimization, which incorporated steepest decent with conjugate gradient methods. Figure 2 shows the geometry optimized structure of the SEBS and SP3 unit cells. First, the packing structure of the neat SEBS, neat PP and SP blends unit cell was examined by the radial distribution functions (RDF). Figure 3 shows the radial distribution function, which is calculated for all atoms of the neat SEBS unit cell. We observed several well pronounced peaks in the range r < 5Å and these peaks illustrate the structure of the SEBS unit cell. The first peak at around 1.07Å corresponds to the bond distance between H and other atoms. The second set peaks at around 1.37 Å and 1.55 Å corresponds to the distance between bonded carbon atoms and non-bonded carbon atoms. Similarly, other molecular peaks such as hydrogen and carbon in H−C−C sequences (2.16 Å), carbon atoms in C−C−C sequences (2.44 Å) were noticed in the RDF of SEBS. No sharp peaks were observed at distances greater than 4 Å and the RDF tend to 1, which is generally considered as the evidence of the amorphous nature of the SEBS unit cell. Figure 4 (left) shows mean square displacement (MSD) of neat SEBS, neat PP and their blends at 150K. The MSD is useful to know the mobility of polymer chains during glass transition process. The slope of the MSD above $T_g$ is much higher than that below $T_g$ as shown in Figure 4 (right). For real measurement of the $T_g$ of polymers using DMA and DSC, the cooling rate has the significant effect. In MD simulations, enough time should be given to the system to reach its minimum energy state. Here in MD simulations, the temperature was set from 500 K to 150 K. The plot of the potential energy (right) and kinetic energy (left) as functions of simulation time (Figure 5a, b) for the model systems at different temperatures indicates that the variation occurs on the well-defined average point and these model systems have been appropriately equilibrated. Figure 5c shows the variation of the bond energy, potential energy and kinetic energy versus temperature for the LAMMPS simulation of SEBS. In all simulations, it was observed that the graph of bond energy, potential energy and kinetic energy of the system increased regularly with increasing the temperature. Additionally, for the proper determination of the glass transition temperature of neat polymers as well as blends, specific volume, density and free volume were also plotted that was obtained from constant-NPT-MD simulation.

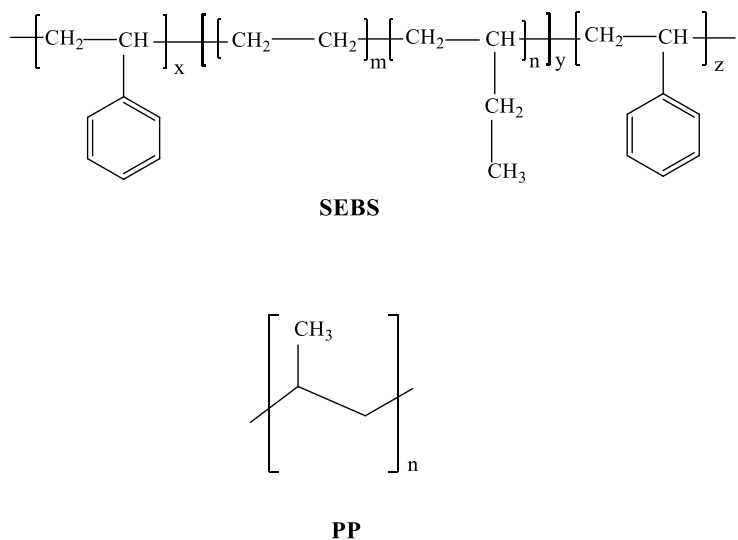

Figure 1. Molecular structure of SEBS and PP.

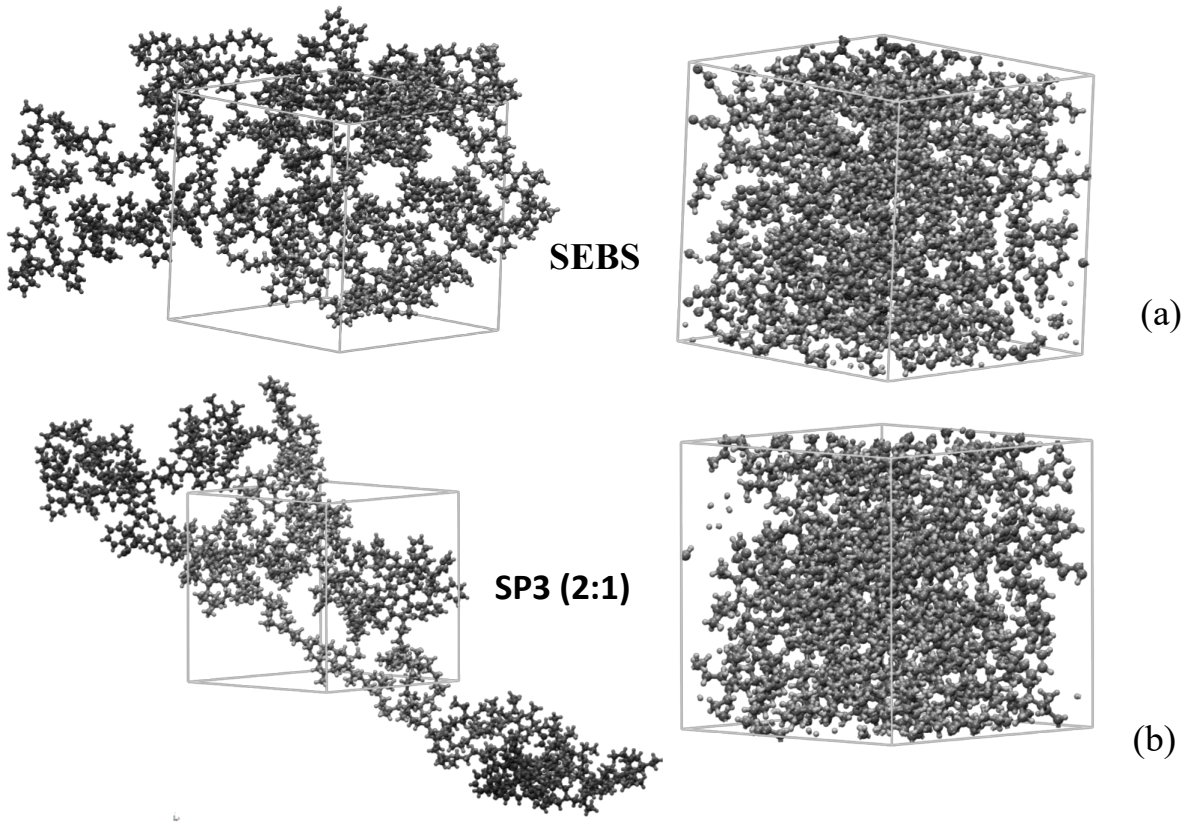

Figure 2. Unit cells of (a) neat SEBS and (b) SP3 (2:1).

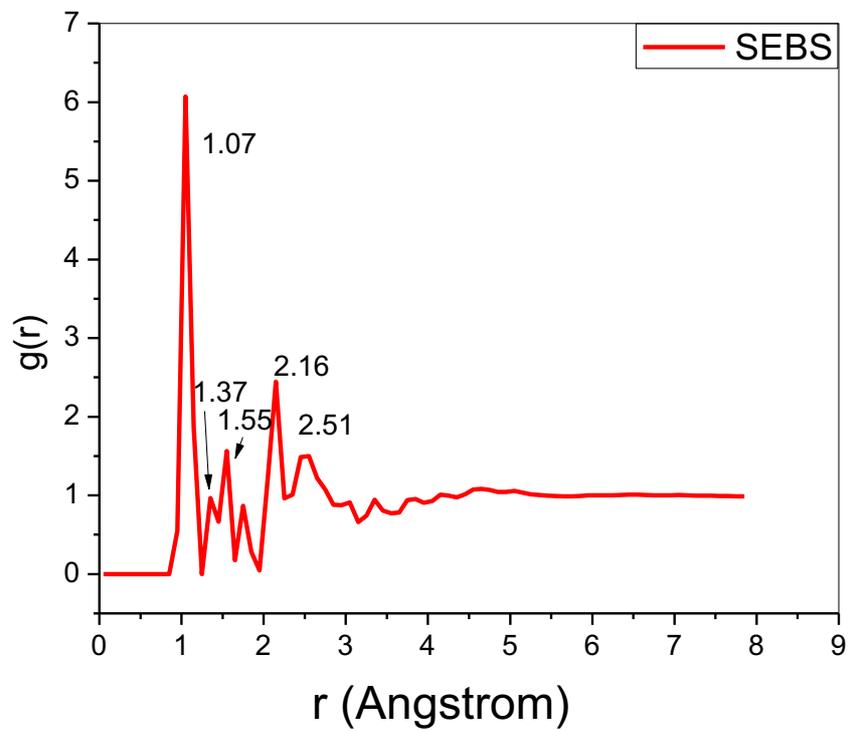

Figure 3. The radial distribution function of the SEBS unit.

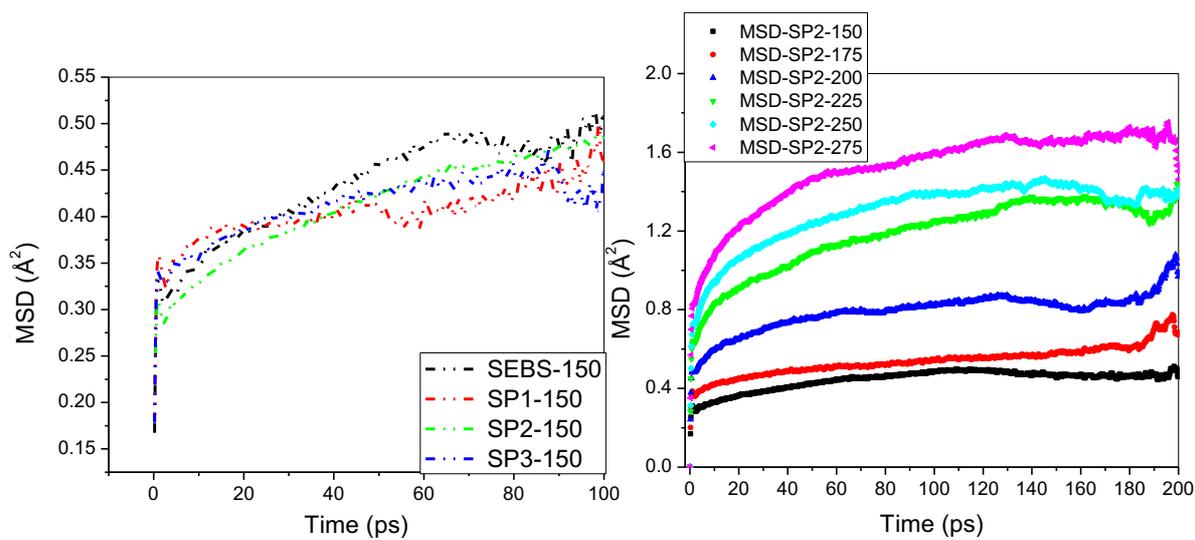

Figure 4. Mean squared displacements (MSD) curves of SEBS, SP1, SP2 and SP3 at 150°C (left). MSD curves of SP2 at 150°C, 175°C, 200°C, 225°C, 250°C, and 275 °C (right).

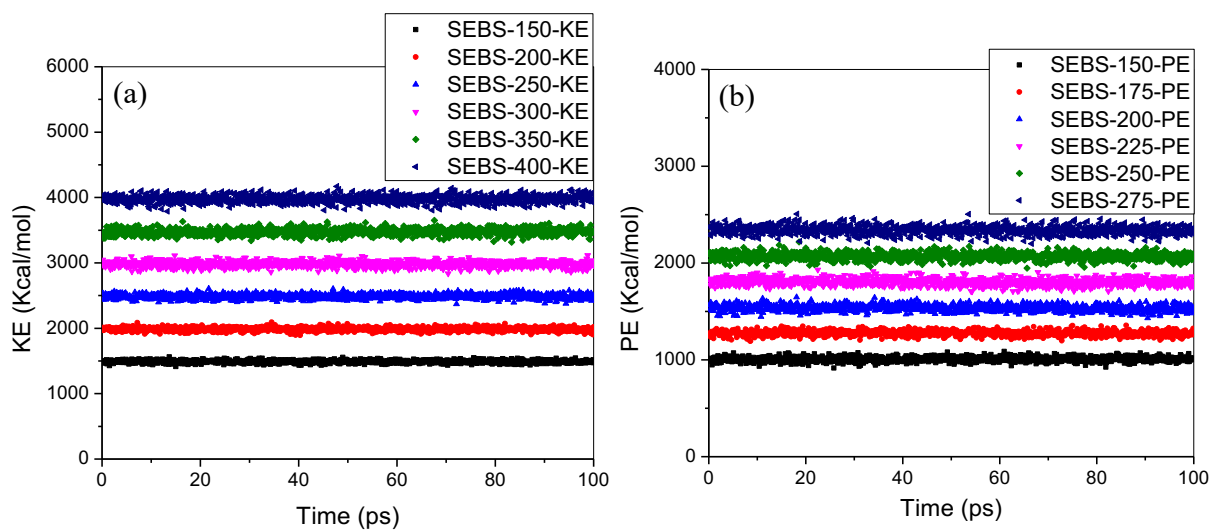

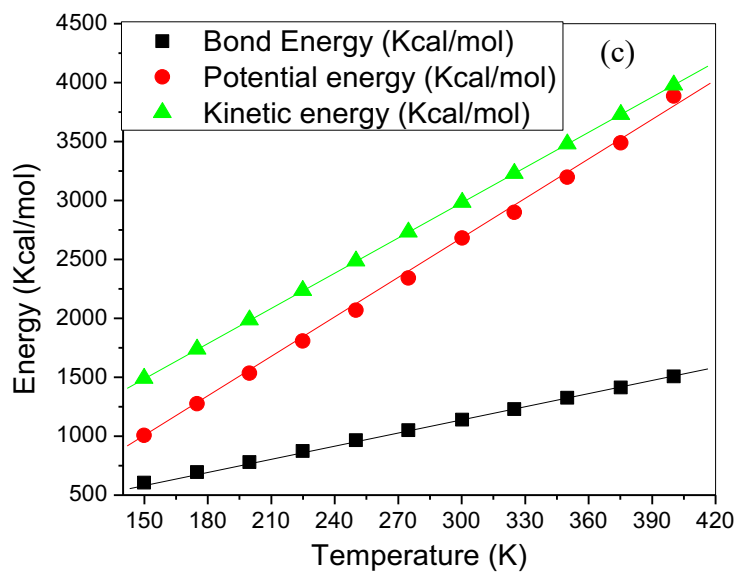

Figure 5. Variation of system kinetic energy (a), potential energy (b) versus time at different temperatures for neat SEBS. Kinetic energy, potential energy and bond energy versus temperature for neat SEBS (c).

## 3. Results and Discussions

*3.1 Thermomechanical properties of SEBS/PP blends*

All blends were prepared using identical two-roll mixing and compression molding conditions and we believe that the blends studies of SEBS and PP brings the interesting characteristics in thermomechanical properties. The stiffness of the material is evaluated based on the value of E'. Figure 6 shows the effects of temperature on the storage modulus (E') of SEBS and PP and their blends at different ratio. Based on the value of E', the variation of the stiffness of the material was observed at broad temperature range from -80°C to 160°C. All the E' curves show three distinct regions: a glassy high modulus region, a transition zone, and a rubbery region (a drastic decay in the modulus in the flow region). Typically, SEBS shows a high modulus below its $T_g$ followed by a strong decrease in its magnitude around -41°C. This strong drop in modulus with temperature around -41°C directs distinct transition from glassy to rubbery state. At any fixed rate of deformation, the temperature at which E' starts to decline rapidly relates to the glass transition temperature. But, in the case of PP changes in the E' are less severe around the glass transition zone (around 9°C) because of its semi-crystalline nature. In the semi-crystalline materials like PP and HDPE, the crystalline chains are arranged in a regular order, and they will remain intact until the temperature reaches the melting point ($T_m$). Only the amorphous part of semi-crystalline polymer experiences segmental motion [33]. Thus, in the case of PP, the E' decreases to a smaller extent than SEBS at this transition zone. Furthermore, neat PP shows the highest modulus than the SP blends at all temperature range from -80°C to 160°C. A small drop in its magnitude of E' was observed at ~9°C due to only the amorphous part of PP undergoes segmental motion. The crystalline chains are arranged in a regular order in the semi-crystalline materials like i-PP and it will stay unbroken until the temperature reaches the melting point ($T_m$). It is worth noting that PP has the maximum and SEBS has the minimum E' values, and the E' values of the blends are found to be intermediate between those of pure components depending on the ratio of SEBS and PP. The value of E' decreases with decrease in the concentration of PP content in SP blend, which is more pronounced above the $T_g$ of the soft segment of SEBS. Relatively, SEBS has very low modulus in the rubbery plateau region and the nature of two-step curves in the SEBS thermograms changes with increasing the PP in SP blend due to good miscibility of SEBS with PP. Specially, hard domain of SEBS is largely affected with the addition of PP as a result the modulus of the blend increases significantly [45–47]. At low concentration of PP in SP blend, PP acts as dispersed phase that will change into co-continuous morphology. Previous study shows that the blend modulus is dominated by the matrix component in the dispersed structures whereas the storage modulus-temperature dependence reflects a greater in co-continuous structures contribution of both components [45].

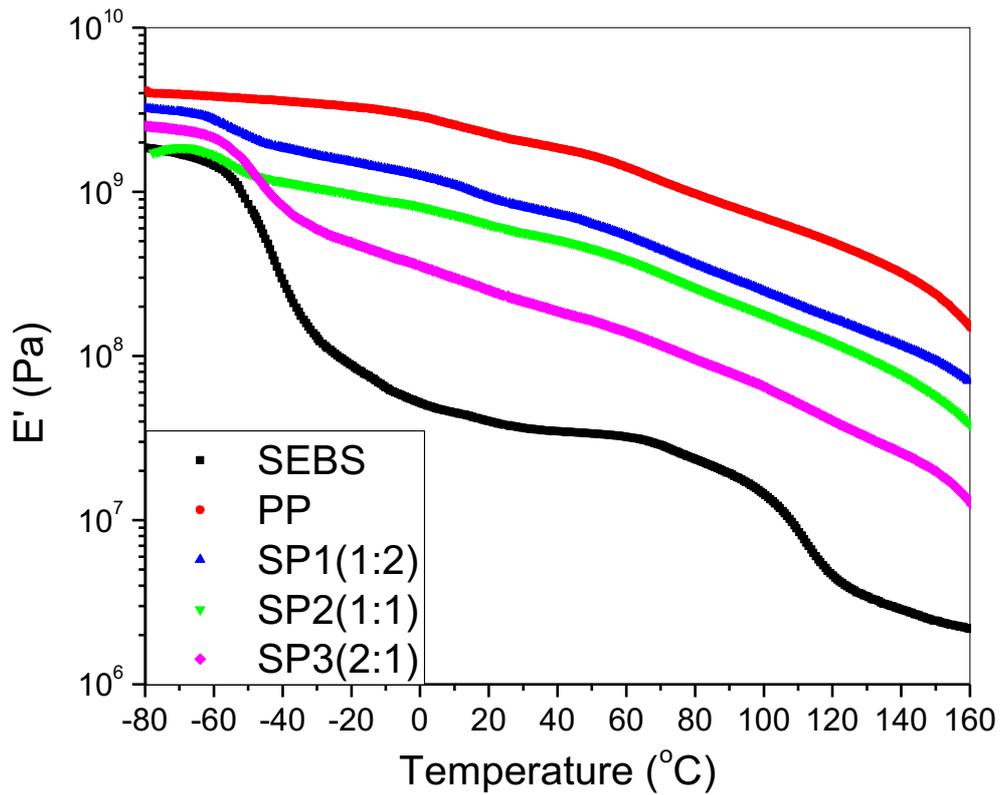

Figure 6. Storage modulus curves of neat SEBS, neat PP, SP1, SP2 and SP3.

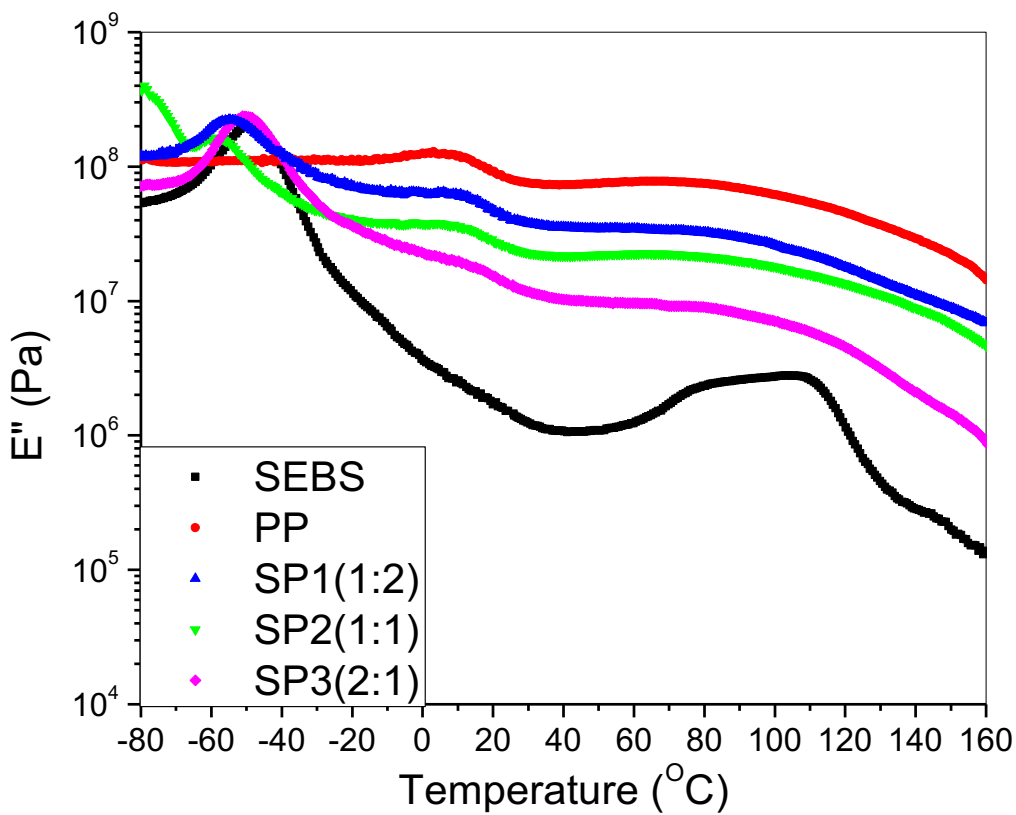

Figure 7. Loss modulus curves of neat SEBS, neat PP, SP1, SP2 and SP3.

Figure 7 shows variation of loss modulus (E'') with temperature of SEBS, PP and SP blends. The loss modulus peak corresponds to the maximum heat dissipation per unit deformation. The peak position of E'' is considered as glass transition temperature when plotted as a function of temperature. From the figure 7, SEBS shows a peak around -41°C corresponding to the $T_g$ of soft segment. Similarly, PP exhibits a weak broad peak around 15°C. In SP blends, one distinct peak along with another weak broad peak indicate the glass transition temperatures the soft segment of SEBS and PP, respectively. The less compatibility between the phases occurred at higher concentration of PP in SP blend. Here, we found, increased loss modulus with increasing PP content in SP blend. In the previous study, some researchers reported that for compatible polymer blend, a single peak is found for the combined processes [48]. For the partially compatible systems, broadening of the transition occurs. Shift in the $T_g$ to higher or lower temperatures as a function of composition also indicates the partial miscibility. Though a modulus determines hardness or stiffness of a material, E'' of polymer indicates energy dissipation. So, E'' must be interpreted as force/energy dissipation. Upon heating, both E' and E'' decrease because less force is required for deformation. In the region of the $T_g$, molecular segmental motions are activated, however motions occur with difficulty. This situation is described as molecular friction that dissipates much of the energy/force as heat and the loss modulus increased [37]. Furthermore, much less energy is stored since the molecules can move with the force giving a rapid decline in E'. With increasing temperature above the $T_g$, molecular frictions are reduced, and less energy is dissipated as a result the E''again decreases.

Dynamical mechanical analysis has been used to predict the miscibility of polymeric systems. Generally, for incompatible polymer blend, the glass transition temperatures of individual polymers can see clearly in the tanδ versus temperature curve. A single peak or probably a new peak is observed when the blend components are compatible for the combined processes. For the partially compatible systems, broadening of the glass transition occurs. The partial miscibility of blend is observed by shift the $T_g$ to higher or lower temperatures as a function of composition. The variation of loss tangent (Tanδ) with temperature for the neat SEBS, PP and SP blends is shown in Figure 8, where SEBS showed $T_g$ at two different temperature corresponding to the soft segment $T_g$ ($T_{g1}$) around -41°C and hard segment $T_g$ ($T_{g2}$) around 113°C. Neat PP shows relatively low intensity Tanδ peak (around 9°C) than SEBS. The loss tangent (Tanδ) goes through a maximum near $T_g$ in the transition region, and then a minimum in the rubbery region. Below $T_g$, the damping is low due to insufficient thermal energy to cause rotational and transnational motions of the segments [33] as a result the chain segments are frozen and viscous flow is low. Furthermore, both frozen in (below $T_g$) or free to move (above $T_g$), damping is low. SEBS have amorphous nature with only physical cross-links between hard and soft segments, and the uncoiling and recoiling process occurs above soft segment $T_g$. On the application and removal of the stress generates more permanent deformation and thus recording the highest loss tangent (tanδ max) values. The SEBS has the highest two damping peaks due to soft and hard segment transition. It is known that higher the tanδ max, the larger the mechanical losses. The results also recommended that increasing the SEBS content caused an increase of elastic behaviour. For neat PP, the damping behaviour is lower than SEBS. So, SEBS display maximum value of tanδ indicating their excellent damping behaviour than PP. The variation of loss tangent (Tanδ) with temperature for the SP blends shows $T_g$ at three different temperature corresponding to the soft segment Tg ($T_{g1}$), hard segment $T_g$ ($T_{g2}$) and $T_g$ of the PP. Generally, blending of elastomer like SEBS with the thermoplastic like PP, shifting of the hard and soft segments peaks occurs along with the peak of PP at high concentration of PP in SP blends. With the variation of PP at the fixed mass of SEBS did not show significant changes on the $T_g$ contributed by PP of SP bend, but the $T_g$ of the soft segment $T_g$ ($T_{g2}$) is significantly shifted to the lower temperature. Decreasing the amount of PP, the $T_g$ of PP in SP blend was disappeared. Furthermore, the amount of PP significantly affected the $T_g$ of hard segment of SEBS in SP as a result the peaks becomes broader and completely disappear. Moreover, the soft segment $T_g$ of SP blend has prominent effect at higher concentration of PP due to the formation of co-continuous phases of PP and SEBS. However, at low concentration of PP in SP blend, the soft segment $T_g$ of blends is nearly close to the soft segment $T_g$ of neat SEBS. It is worth noting that the PP largely affects the hard segment $T_g$ of SEBS. Hard segment $T_g$ of SEBS completely disappeared at more than 40 wt% of PP in SP blend. At the higher concentration of PP in SP blend, three distinct and clearly separate tanδ peaks corresponding to the $T_g$'s of SEBS and PP indicating that the blends have two co-continuous phases and are incompatible. On the contrary, The $T_g$ of PP disappeared in SP blend at low concentration of PP. The $T_g$ of neat SEBS and PP as well as blends obtained from the Tanδ curves are given in Table 1. In the SP blends, the crystalline plastic phase acting as physical crosslinks imposes some restriction towards cyclic loading and the Tanδ$_{max}$ decreases with increases in PP content.

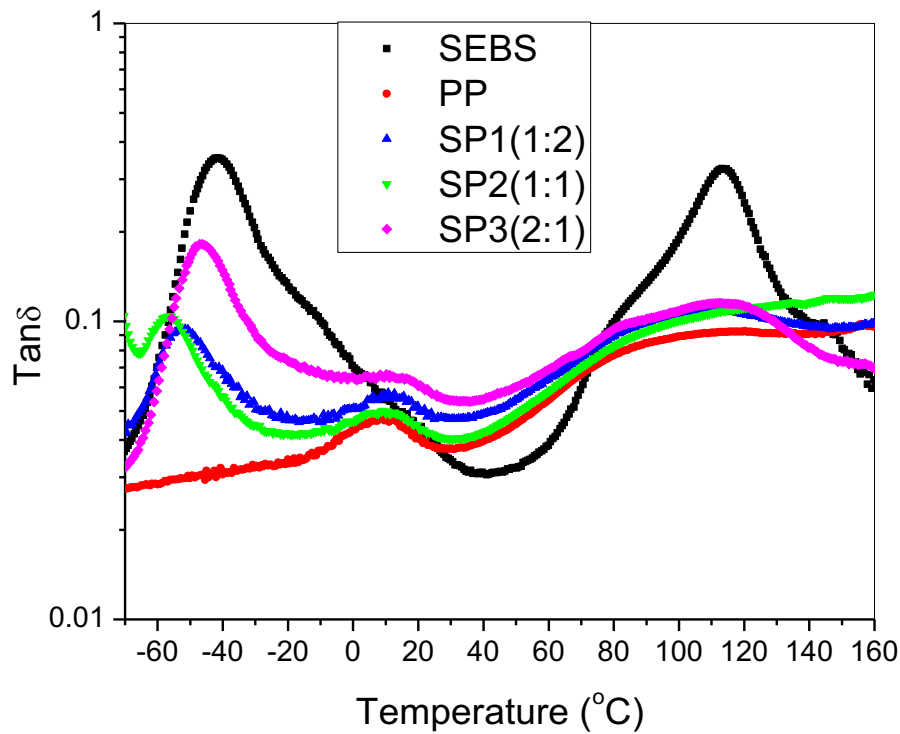

Figure 8. Tanδ curves of neat SEBS, neat PP, SP1, SP2 and SP3.

*3.2 Evaluation of glass transition temperature in LAMMPS*

To illustrate glass transition phenomenon at the molecular scale, we applied Large-scale Atomic/Molecular Massively Parallel Simulator (LAMMPS). The total potential energy ($E_{total}$) was expressed as the sum of the valence terms ($E_{bond}$) and non-bond interaction terms ($E_{non-bond}$) [37].

$$E_{total} = E_{bond} + E_{non-bond} = E_b + E_\theta + E_\varphi + E_{inv} + E_{elec} + E_{vdW} \quad \ldots\ldots\ldots\ldots\ldots\ldots(11)$$

Here, $E_b$, $E_\theta$, $E_\phi$ and $E_{inv}$ represent bond stretching, angle bending, dihedral torsion and inversion (out-of-plane) in the valence terms. On the other hand, $E_{vdW}$ and $E_{elec}$ are used for calculating van der Waals force and electrostatic interactions, respectively. For the calculation of the non-bond interactions, cut-off being set to 0.95 nm, which is less than half of the cell length (during the glass transition process, cell length was kept around 2.2 – 2.5 nm). Theodorou and Suter's method [49–51] was used to build the amorphous polymer models of neat PP and neat SEBS as well as their blends at different ratio. In our study, each cell consisted of a parent polymer chain with 100 repeat units. Simulation was conducted from 500 K to 150 K. At each temperature, constant-NVT MD simulation was conducted for 30 ps. Then, molecular dynamics simulations of neat PP, neat SEBS, and blends of SEBS and PP are performed in the NPT ensemble. The densities of PP and SEBS are 0.946 g/cm$^3$ and 0.94 g/cm$^3$, respectively are used for LAMMPS simulation. The $T_g$ prediction was compared with experimental results of neat polymers SEBS, PP as well as SP blends, where the predicted $T_g$ was found equivalent to our experimental value of $T_g$.

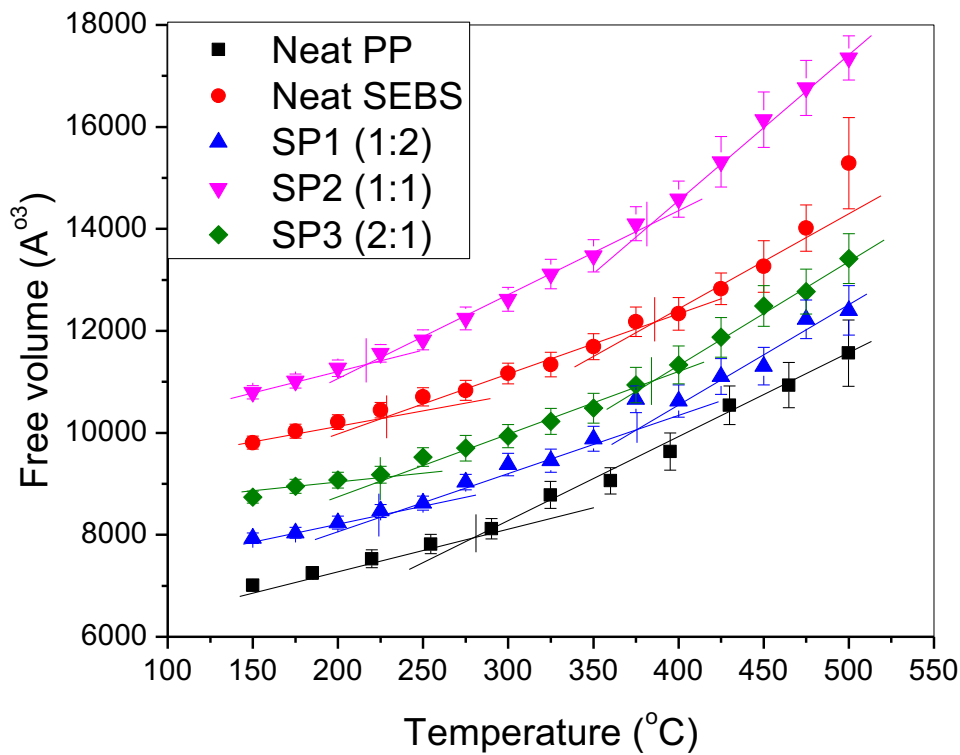

Figure 9. The free volume-temperature curves of the neat SEBS, neat PP, SP1, SP2 and SP3.

*Free volume*

Based on the Fox and Flory's theory of glass transition, an abrupt change on the free volume of the polymer occurs during the cooling down, and that point is called the glass-transition temperature. During the glass transition process, free volume defines how much space can be provided for the segment motion of a polymer chain. Figure 9 displays the plot of free volume of neat SEBS and PP as well as their blends in which the $T_{g1}$ of SEBS, PP, SP1, SP2 and SP3 are observed 229.8K, 282.1K, 225.0K, 215.1K, and 225.6 K, respectively. The $T_{g1}$ from MD simulation showed close agreement with the experimental $T_{g1}$ using dynamic mechanical analysis (Table 1). It is quite interesting to note that we did not observed $T_{g2}$ for SEBS and PP blend at 1:1 and 1:2 ratio using DMA test, but second abrupt free volume change i.e. $T_{g2}$ were observed for them at 380.0°C and 376.5°C, respectively by MD simulation. So, we can confirm that MD simulation is more appropriate to determine the $T_g$ of the complex polymer blends.

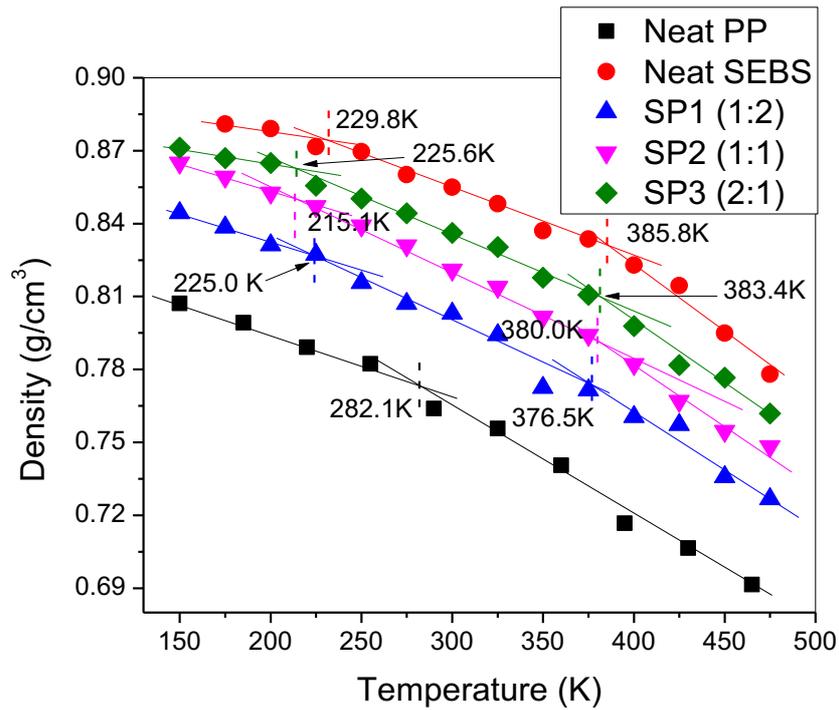

Figure 10. The density-temperature curves of the neat SEBS, neat PP, SP1, SP2 and SP3.

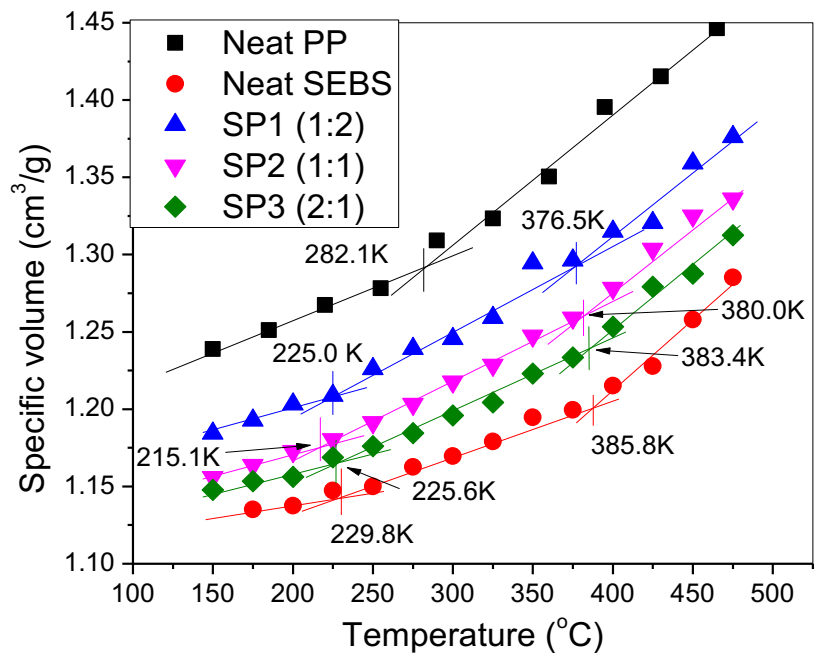

Figure 11. The specific volume-temperature curve of the neat SEBS, neat PP, SP1, SP2 and SP3.

*Density and specific volume*

In this study, we specially focused on the effect of PP on the $T_g$ of soft segment and hard segment of SEBS during the preparation of SP blends with the variation of PP content. Figure 10 shows the density verses temperature of SEBS, PP, SP1, SP2 and SP3. The dynamic change of density of neat polymers and blends shows yielding a kink in the curve at $T_{g1}$ and $T_{g2}$ which is useful to get the comparable $T_g$ from experimental study. We observed significant effect of the soft segment $T_g$ of SEBS with the addition of PP in SP blends towards lower temperature as observed in the experimental results (Table 1). In SEBS, the soft segment forms a continuous phase and that determines the elasticity of the material, where hard segment regulates the rigidity. Generally, the percentage of the hard segment and soft segment in SEBS varied based on the desired application of SEBS [45,46]. For the commercial applications with enough elasticity, SEBS contains ~30% hard segment. However, by MD simulation, we clearly observed $T_{g2}$ of the hard segment of SEBS at 385.8K. Furthermore, our experimental results from DMA shows a hard segment $T_{g2}$ of SEBS around 386.2K, which is very close to the predicted hard segment $T_{g2}$ of SEBS by MD simulation. Moreover, when the blend ratio of SEBS and PP varied from 1:1, 1:2, and 2:1, we observed $T_{g1}$ of the soft segment as well as $T_{g2}$ of hard segment of SEBS which are close to experimental results. Interestingly, we obtained extra information about the $T_{g2}$ of SP blend at 1:2 and 1:1 ratio that was not possible even by DMA test. These results indicated that the $T_{g2}$ of the hard segment as well as $T_{g1}$ of the soft segment of SEBS can be precisely visualized by MD simulation even blend with PP. Figure 11 shows the temperature dependence of the specific volume of neat SEBS, neat PP and their blends at three different ratio. Below $T_g$, the increment of specific volume is less pronounced with increasing temperature as a result yielding a kink in the curve which is useful to determine $T_g$ [52]. The $T_{g1}$ of neat SEBS, neat PP, SP1, SP2 and SP3 are observed 229.8K, 282.1K, 225.0K, 215.1K, and 225.6 K, respectively. All the SP blends have lower $T_{g1}$ than SEBS and PP. Similarly, our predicted $T_{g2}$ of the hard segment of SEBS by LAMMPS even in blend system showed close agreement with the experimental results. We found consistent simulation result with our experimental results as shown in Table 1. Therefore, we confirmed that MD simulation is an effective method to predict $T_g$ of complex polymer blends.

## 4. Conclusions

Molecular simulations for the prediction of thermal events in neat SEBS, neat PP and their blends SP1, SP2 and SP3 are shown good agreement with experimental DMA results for thermomechanical properties. Here, the storage moduli of SP blends were found to be intermediate between the pure components; SEBS and PP. As the concentration of the PP increases the storage modulus and loss modulus also increase in SP blends. The analysis of density, specific volume, free volume obtained from the LAMMPS simulation showed $T_{g1}$ 229.8K, 282.1K, 225.0K, 215.1K, and 225.6 K for neat SEBS, neat PP, SP1, SP2 and SP3, respectively. Similarly, the hard segment $T_{g2}$ of SEBS and SP3 by simulation were observed 385.8K and 383.4K, respectively. These glass transition temperature values showed closed agreement with the experimental $T_g$ obtained from DMA. Interestingly, SP1 and SP2 showed $T_{g2}$ 376.5K and 380.0K, respectively by MD simulation which are not even possible to measure by DMA. This study establishes the possibility of computer aided design and analysis of polymer blends with desired thermomechanical properties.

**Acknowledgements**

We acknowledge the grant support from Higasket Plastic Group Co. Ltd on completing this work.